\documentstyle[preprint,aps,prl]{revtex}
\tighten
\draft
\begin{document}
\preprint{\today}
\title{Cooling rate dependence of the internal structure of a Lennard-Jones
glass}
\author{Katharina Vollmayr\cite{kvollmayr}, Walter Kob\cite{wkob}
and Kurt Binder}
\address{Institut f\"ur Physik, Johannes Gutenberg-Universit\"at,
Staudinger Weg 7, D-55099 Mainz, Germany}
\maketitle

\begin{abstract}
Using molecular dynamics computer simulations we investigate the
dependence of bulk properties of a Lennard-Jones glass on the cooling
rate with which the glass was produced. By studying the clusters
formed by the nearest neighbor shells of the particles, we show
that the internal structure of the glass depends on the cooling rate
and that the cooling rate dependence of the enthalpy of the glass
can be understood by the change of this internal structure.

\end{abstract}

\narrowtext


If a material is cooled fast enough from a high temperature liquid
state to a low temperature solid state, the crystallization of the
substance at or slightly below its melting point can be avoided and the
end product will be an amorphous solid, i.e. a glass. Since this
so-called glass transition is essentially the falling out of
equilibrium of the system because the typical time scale of the
experiment is exceeded by the typical time scale of the relaxation
times of the system, the resulting glass can be expected to depend on
the way the glass was produced, e.g.,  on the cooling rate of the
sample or the particulars of the cooling schedule. Such a dependency of
the density of the glass or of the glass transition temperature has
indeed been found in experiments~\cite{exp_glass_cool,bruning94} and in
computer simulations~\cite{cs_glass_cool}. Although the presence of
such a cooling rate dependence is therefore well established (although,
e.g., Speedy reports that no dependency of the properties of a hard
sphere glass on the quench rate could be found~\cite{speedy}), the
exact reason for this dependence is still not known.  Is it mainly that
the {\it global} arrangement of the particles depends on the cooling
rate, or is this global arrangement essentially cooling rate
independent (of course only in a statistical sense) and it is only that
the positions of the particles {\it within} this arrangement depend on
the cooling rate? E.g., using a molecular dynamics computer simulation
J\'onsson and Andersen have found that a binary Lennard-Jones glass has
at low temperatures clusters of interpenetrating icosahedra that
percolate throughout the sample~\cite{jonsson88}. Thus one might
wonder  whether this whole percolating structure shows a cooling rate
dependence or whether the topology of the structure is basically
independent of the cooling rate and it is only the details of the
positions of the particles {\it within} this structure that give rise
to the cooling rate dependence of the various macroscopic properties of
the glass.  Since computer simulations are a most useful tool to study
the full microscopic details of such systems they should allow to gain
some insight into this question and in this Letter we report some of
our findings of our investigations in this direction.

The model we are considering is a binary mixture of Lennard-Jones
particles (type A and type B), both of them having the same mass $m$.
The interaction between particles of type $\alpha$ and $\beta$
($\alpha,\beta \in \{A,B\}$) is given by
$V_{\alpha\beta}(r)=4\epsilon_{\alpha\beta}[(\sigma_{\alpha\beta}/r)^{12}-
(\sigma_{\alpha\beta}/r)^6]$ with $\epsilon_{AA}=1.0,
\epsilon_{AB}=1.5, \epsilon_{BB}=0.5, \sigma_{AA}=1.0, \sigma_{AB}=0.8$
and $\sigma_{BB}=0.88$. In order to decrease the computational load
$V_{\alpha \beta}(r)$ was truncated and shifted at $r=2.5\sigma_{\alpha
\beta}$. This potential has been used previously to study the dynamics
of strongly supercooled liquids and thus it is known that this system
is a good glass former, i.e. not prone to
crystallization~\cite{kob94}.

In the following all results are given in reduced units with
$\sigma_{AA}$ the unit of length, $\epsilon_{AA}$ the unit of energy
and $(m \sigma_{AA}^2/48 \epsilon_{AA})^{1/2}$ the unit of time.  Using
the constant pressure algorithm by Andersen~\cite{andersen80} the
equations of motion were integrated with a step size of 0.02 and an
external presssure of 1.0. The system was equilibrated at a high
temperature $T_0=2.0$ where it is in a liquid state. Subsequently it
was coupled to a stochastic heat bath whose temperature $T_b$ was
decreased linearly in time, i.e.  $T_b(t)=T_0-\gamma t$, where $\gamma$
is the cooling rate. This quench was done until the temperature of the
heat bath was zero. Using a steepest descent procedure, the resulting
configuration of particles was relaxed to its nearest metastable
minimum in configuration space and the so obtained configurations were
subsequently analyzed. Note that during this steepest descent procedure
also the volume of the system was allowed to relax. The range of
cooling rates we investigated was $3.1\cdot 10^{-6} \leq \gamma \leq
2.0\cdot 10^{-2}$. The number of A and B particles was 800 and 200,
respectively. In order to improve the statistics of our results, we
performed for each value of $\gamma$ ten different runs and averaged
over these runs.

In Fig.~1 we show $H_f(\gamma)$, the final enthalpy of the
system after having quenched the system to $T=0$ with a cooling rate
$\gamma$.  We see that a variation of $\gamma$ over about four decades
leads to a change in $H_f(\gamma)$ of only 1-2\%.  The smallness of
this change shows that in order to investigate such cooling rate
dependencies it is important to vary $\gamma$ over several
orders of magnitude and to use either fairly large systems or to
average over many small ones.\par

Qualitatively the same behavior as in the case $H_f(\gamma)$ was found
also in other quantities such as the density, the coordination number
of the particles or the radial distribution functions. In all cases the
effect of the variation of the cooling rate was to give rise to a
change of a few percent of the quantity investigated~\cite{vollmayr95}.
The size of these cooling rate effects is comparable with the one found
in real experiments, in that, e.g., Br\"uning {\it et al} found that the
density of B$_2$O$_3$ also changes by about 1\% when the cooling rate
is varied over three decades~\cite{bruning94}.

Now that we have demonstrated that our system shows, similar to
experiments, a cooling rate dependence of various bulk properties of
the glass, we can investigate the problem mentioned in the introduction,
namely whether these properties depend on the cooling rate because the
whole internal structure of the glass depends on the cooling rate or
whether this structure is essentially cooling rate independent and the
bulk properties of the system change only because the particles
re-adjust their position by small amounts without that the properties of
this global structure are changed.

In order to address this question we investigated the cooling rate
dependence of the properties of the nearest neighbor shells of the
particles. The nearest neighbor shell of a particle of type $\alpha
\in\{A,B\}$ was defined to be the set of all those particles that have a
distance $r$ which was less than the location of the minimum in the
corresponding radial distribution function $g_{\alpha \beta}(r)$. For
the sake of clarity we will discuss in the following only the nearest
neighbor shells of the A particles, i.e. of the majority species.
However, similar results have also been found for the B particles.

We define $N_{A_{z,\mu}}$ to be the number of A particles that have $z$
nearest neighbors with $\mu$ of them being a B particle. In the
following we will call the collection of particles given by the central
particle and its nearest neighbor shell a ``cluster''. The energy $E_c$
of such a cluster is defined to be the sum over all the interactions of
the particles within this cluster. In Fig.~2 we show the dependence of
$E_c$ on the cooling rate. In order not to crowd the figure too much we
show only the curves for a subset of the clusters we found, but the
curves for the clusters not displayed show qualitatively the same
$\gamma$ dependence as those shown. Three observations can be made from
this figure: 1) The energy of all clusters is decreasing with
decreasing $\gamma$. 2) For every given type of cluster the difference
in $E_c$ between the fastest and the slowest cooling rate is on the
order of 1\%. 3) For a given cooling rate the difference between the
energy of the various clusters is generally much larger than 1\%.

Since the enthalpy of the system is related to the weighted sum of the
energy of the clusters and since we have seen in Fig.~1 that the former
shows a change of about 1\% when the cooling rate is varied over
several decades, one might conclude from points 1) and 2) that the
cooling rate dependence of the enthalpy of the system is given entirely
by the cooling rate dependence of the clusters. However, before this
conclusion can be drawn it is of course necessary to investigate
whether the frequency of the clusters of the various types does not
show a significant $\gamma$ dependence, since the third point shows
that such an effect might be important.

In Fig.~3 we show $P_{A_{z,\mu}}$, the probability to find a cluster of
type $A_{z,\mu}$, for all those combinations of $z$ and $\mu$ for which
this probability is not too small. From this figure we recognize that
there are clusters whose corresponding $P_{A_{z,\mu}}$ is essentially
independent of the cooling rate, such as $A_{13,2}$, and that there are
other clusters, marked with symbols, whose $P_{A_{z,\mu}}$ changes
significantly with $\gamma$. Among these latter types of clusters there
are some whose $P_{A_{z,\mu}}$ changes by as much as 20\% (e.g.
$A_{12,1}$).  Thus if one takes into account that the difference
between the energy between the various clusters is usually
significantly larger than 1\% (see point 3) and Fig.~2) such a change
in $P_{A_{z,\mu}}$ can lead to a change in the enthalpy of the system
that is significantly larger than the one that comes from the change of
the energy of the cluster itself.

In order to demonstrate this effect clearer we show in Fig.~4
$P(E_{c})$, the probability that a cluster has an energy $E_{c}$,
versus $E_{c}$ for all cooling rates investigated. We see that this
distribution function shows various peaks, each of which can be
identified with particular types of clusters. From this figure we also
recognize that the main difference between the distribution for the
fastest cooling rate (bold dashed curve) and the distribution for the
slowest cooling rate (bold solid curve) is that the {\it height} of
these peaks changes. Since the position of the peaks is given by the
mean energy of the corresponding cluster and since we find that this
position is essentially independent of $\gamma$ we thus come to the
conclusion that the enthalpy of the system shows a cooling rate
dependence because the frequency of the various clusters shows a
strong dependence on $\gamma$ and {\it not} because the energy of the
individual clusters is changing.

In this Letter we have focused on investigating the cooling rate
dependence of the enthalpy of the system. Very similar results have
also been obtained for the cooling rate dependence of the
density~\cite{vollmayr95}. Also for this quantity we found that it is
the $\gamma$ dependence of the distribution function for the clusters
rather than the individual properties of the clusters themselves that
gives rise to the $\gamma$ dependence of the density. Thus we can
summarize this work by saying that the cooling rate effects observed in
structural glass formers follow from the fact that the internal
structure of the glass depends on the cooling rate and not that
there exists an internal structure that is independent of the cooling
rate and that the particles within this structure show a cooling rate
dependence.  Although our findings apply, strictly speaking, only to
the system investigated here, it can be expected that the same or at
least a similar mechanism is present also in other structural glass
formers and thus our conclusions should be valid for a relatively large
class of glass formers.

Acknowledgements: We thank C. A. Angell for valuable discussions. K. V.
thanks Schott Glaswerke, Mainz, for financial support. Part of this
work was done on the computer facilities of the Regionales
Rechenzentrum Kaisers\-lautern.

\newpage
FIGURES\\
{Fig. 1: Final enthalpy $H_f$ versus the cooling rate $\gamma$.
\label{fig1}}\\
{Fig. 2: $E_c$, the energy of a cluster, for a few selected types of
clusters around A particles versus the cooling rate $\gamma$. See
text for the meaning of the two indices identifying the type of the
cluster.
\label{fig2}}\\
{Fig. 3: $P_{A_{z,\mu}}$, the probability to find a cluster of type
$A_{z,\mu}$, versus the cooling rate $\gamma$. Only the most frequent
types of clusters are shown. See text for the meaning of the two
indices identifying the type of the cluster.
\label{fig3}}\\
{Fig. 4: $P(E_c)$, the probability that a cluster has an energy $E_c$,
versus $E_c$ for all cooling rates investigated. The bold solid and
dashed lines correspond to the slowest and fastest cooling rates,
respectively.
\label{fig4}}\\
\end{document}